\documentstyle[twocolumn,aps]{revtex}

\input epsf

\begin{document}
\title{Inelastic Collisions of Ultracold Polar Molecules}
\author{John L. Bohn \cite{byline}}
\address{JILA, National Institute of Standards and Technology and
University of Colorado, Boulder, CO 80309-0440}
\date{\today}
\maketitle

\begin{abstract}
The collisional stability of ultracold polar molecules in
electrostatic traps is considered.  Rate constants for collisions
that drive molecules from weak-field-seeking to strong-field-seeking
states are estimated using a simple model.
The rates are found to be quite large, of order $10^{-12}$-$10^{-10}$
cm$^3$/sec, and moreover to grow rapidly in an externally applied
electric field.  It is argued that these results are generic
for polar molecules, and that therefore polar molecules should
be trapped  by other than electrostatic means.
\end{abstract}


\narrowtext

Recently Bethlem {\it et al.} broadened the scope of ultracold AMO
physics by cooling and electrostatically trapping ND$_3$ molecules
\cite{Meijer1,Meijer2}.  This achievement is notable both for the
complexity of the species trapped and for the generality of the
Stark slowing technique, which could in principle cool any polar molecule.
This technique is therefore now on a par with other experimental methods
for producing cold molecular gases, such as photoassociation \cite{PA}
and buffer-gas cooling \cite{BG}, as well as photoproduction of
molecular ions in a Paul trap \cite{Molhave}.

The electrostatic trap demonstrated in Ref. \cite{Meijer2} raises anew 
questions of collisional stability that are  familiar in the context of 
magnetic trapping of atoms \cite{trap}.  An electrostatic trap
can only confine dipoles that are in their weak-field seeking states,
since Maxwell's equations permit a local field minimum but not
a field maximum.  Thus the dipoles are succeptible to orientation-changing
collisions that populate the strong-field seeking, untrapped states.
In the case of magnetic trapping of alkali atoms a standard remedy
against collisional losses is to prepare the atoms in their stretched
spin states, whereby the dominant spin-exchange collisional processes are
absent.  Atomic spins can then only change their orientation via
spin-spin dipolar processes, which are weak because of the inherent
weakness of magnetic dipolar interactions. 

The purpose of this Rapid Communication is to point out that polar molecules 
are not as immune to dipolar relaxation as are magnetic atoms, simply
because electric dipoles have a much stronger interaction.  Indeed, the 
force between a pair of $d=1$ Debye (0.39 atomic units) electric dipoles is
$\sim 3 \times 10^3$ times larger than that between a
pair of $\mu = 1\mu _{B}$ magnetic dipoles.  
The basic physics of electric dipolar relaxation lies in the competition
between the dipoles' interaction with the electric field when they are far apart,
$-{\vec d} \cdot {\vec {\cal E}}$, and with each other when they
are closer together, $\sim {\vec d}_1 \cdot {\vec d}_2 / R^3$.  At small 
values of intermolecular separation $R$ the dipoles will tend to lock on their
intermolecular axis rather than on the lab-fixed axis set by 
${\vec {\cal E}}$; the competition between these tendencies 
scrambles the orientation of the molecular dipoles.
The resulting state-changing collisions
can in principle be suppressed by a field strong enough to maintain
the dipolar orientation.  This will happen, roughly, if 
$d {\cal E} > d^2/R^3$ for small values of $R$.  
Still, for $d=1$ Debye dipoles at a typical collision distance
$R \approx 10$ atomic units (a.u.), an electric field of
$10^6$ V/cm would be required to maintain dipolar orientation.  Thus
very large laboratory fields may exert some mitigating influence,
but are unlikely to arrest relaxation altogether.

To quantify this general argument this paper presents detailed calculations 
using a simplified  model of collisions.  In general the physics of cold
molecular collisions will be quite complex, intertwining rotational,
electronic, nuclear spin, and perhaps even vibrational degrees of
freedom.  However, to establish orders of magnitude for dipolar
relaxation rate constants it suffices to focus on orientational
degrees of freedom, and to account only for the dominant dipole-dipole 
interaction between molecules.  Accordingly, a simplified ``toy''
model is used here, which has zero spin and nuclear spin.
The molecules are assumed to be diatomic rigid rotors
with electric dipole moments $d=1$ Debye along their molecular axes.
The electronic ground state of the molecules is assumed to be
$^1\Pi$, so that it possesses a $\Lambda$-doublet of parity eigenstates.
The splitting of this doublet is assumed to have a ``typical'' value
of $\Delta = 10^{-3}$ cm$^{-1}$, and the lower-energy state is assumed
to have even parity.

At ultracold temperatures and in zero electric field the molecules
occupy parity eigenstates, hence exhibit no permanent dipole moment.
The dipole moments only become apparent when the field is large
enough to significantly mix states of different parity, thus
``activating'' the dipoles.  This occurs at a field value
where the Stark effect transforms from quadratic to linear,
at $\approx 100$ V/cm in the present model (Figure 1).  For fields below
this value the molecules are fairly weakly interacting, whereas
above this value the molecules have extremely strong dipole
couplings.  Thus in the $\Lambda$-doubled state collisions can
be manipulated using modest electric fields, in contrast to the
$\sim 10^{5}$ V/cm fields required to influence cold atomic
collisions \cite{You}.  These arguments also apply to
molecules with $^{2S+1}\Sigma$ electronic symmetry when $S>0$ and the
molecule exhibits an $\Omega$-doubling \cite{DeMille}, as
well as to ND$_3$.

Figure 1 shows the electric field dependence of the lowest-lying energy levels in the
model molecules.  Although both $N=1$ and $N=2$ rotational levels are shown,
the calculations below  focus exclusively on the $N=1$ levels.  Their low-${\cal E}$
behavior is shown in the inset, labeled by the pair of quantum numbers
$|M_N|,p$.  Here $p$ stands for the parity in zero field, while $|M_N|$ 
denotes the magnitude of the molecular rotation's magnetic quantum number
referred to the laboratory axis; the $M_N=1$ and $M_N=-1$ levels are
degenerate even in an electric field.  In an electrostatic
trap of the kind used by Bethlem {\it et al.} the trapped states are
the weak-field seekers, i.e., those whose Stark energy rises with 
rising field.  These are the $||M_N|,p \rangle =$ $|1,- \rangle$ states in the model.

The Hamiltonian for collisions between the molecules consists of
four terms in this model,
\begin{equation}
{\hat H} = {\hat T} + {\hat H}_{\rm fs} + {\hat H}_{\rm field}
+{\hat H}_{\rm dip-dip}.
\end{equation}
Here ${\hat T}$ represents the kinetic energy, ${\hat H}_{\rm fs}$
the molecular fine structure including the $\Lambda$-doubling,
and the last two terms are the electric field interaction and
the dipole-dipole interaction between molecules:
\begin{equation}
{\hat H}_{\rm field} = -({\vec d}_1 + {\vec d}_2) \cdot {\vec {\cal E}},
\end{equation}
\begin{equation}
{\hat H}_{\rm dip-dip} = 
{{\vec d}_1 \cdot {\vec d}_2 - 3({\hat R}\cdot{\vec d}_1)
  ({\hat R}\cdot{\vec d}_2) \over  R^3},
\end{equation}
where ${\hat R}$ denotes the orientation of the
vector joining the centers-of-mass of the molecules.
Dispersion and exchange potentials are neglected here since they are of
secondary importance to dipolar interactions at large $R$.  To avoid
problems with the singularity of $1/R^3$ at $R=0$, vanishing
boundary conditions are imposed at a cutoff radius $R_0 = 10$ a.u.,
where the potentials are deep compared to $\Delta$. 

In the scattering calculation the molecules are assumed to be identical bosons,
so that only even partial waves are relevant.  Only the partial
waves $L=0$ and $L=2$ are included explicitly, even though in
principle all partial waves are coupled together by strong
anisotropic interactions.  However, the neglected higher-$L$
partial waves can be shown, in the Born approximation, to fall
off rapidly with $L$ \cite{Yi,Bohnfuture}.  Cross sections
for processes that change molecular channel $|i \rangle$ into
channel $|i^{\prime} \rangle$ are given as in Ref. \cite{Bohnmol}:
\begin{equation}
\sigma_{i \rightarrow i^{\prime}} = {\pi \over k_i^2}
\sum_{LM_L L^{\prime} M_L^{\prime}}
|\langle i,LM_L | T | i^{\prime} L^{\prime} M_L^{\prime} \rangle |^2,
\end{equation}
where $T$ is the T-matrix for scattering and $k_i$ is the
incident wave number.  Because of the incoherent sum in (4),
contributions arising from $L=0$ and $L=2$ incident partial
waves can be presented individually.  All results, for both elastic
and state-changing collisions will be reported as
event rate coefficients,
\begin{equation}
K_{i \rightarrow i^{\prime}} = v_i \sigma_{i \rightarrow i^{\prime}},
\end{equation}
where $v_i$ is the incident relative velocity of the collision partners.

Although state-changing collisions are primarily of interest here,
it is useful first to consider elastic scattering of
polar molecules, to illustrate the enormous influence of the electric field.
Figure 2 shows the elastic scattering rate constant $K_{\rm el}$ for
molecules in their $|M_N,p \rangle$ $=|1,- \rangle$
state.  Figure 2(a) depicts the low-field limit (${\cal E}=0$ V/cm), while 
Fig. 2(b) shows a higher-field limit (${\cal E}=200$ V/cm), where the
Stark effect is linear.  These rate constants are separated into
their s-wave (solid line) and d-wave (dashed line) contributions.

In both the low- and high-field regimes the s-wave rate has the familiar
$K_{\rm el} \propto E^{1/2}$ threshold behavior as a function of collision energy 
$E$, arising from the Wigner
threshold law.  For the d-waves, however, the threshold dependence
changes dramatically, falling off as $K_{\rm el} \propto E^{5/2}$ at low
field but as $K_{\rm el} \propto E^{1/2}$ at higher fields.  This behavior has to do
with qualitative differences in the long-range intermolecular
potentials.  In the low-field limit, the asymptotic molecular
states are parity eigenstates, hence have no dipole moment.
At large values of $R$, where the dipole-dipole interaction
energy becomes less than the $\Lambda$-doubling energy, the
dipole-dipole $1/R^3$ potential is thus effectively absent.
By contrast, in the high-field
regime parity eigenstates are mixed at large $R$ and the $1/R^3$
potential is activated.  It is well known \cite{Shakeshaft,Sadeghpour}
that a $1/R^3$ potential contributes a long-range scattering phase shift 
that is proportional to the wave number $k$ for {\it all} partial
waves $L>0$, thus yielding a $K_{\rm el} \propto E^{1/2}$ threshold law.  

This ``switching on'' of the dipolar interaction in the presence
of a field also manifests itself in the inelastic rate constants,
which are shown in Figure 3.  This figure shows the sum 
$K_{\rm relax} = \sum_{i^{\prime}} K_{i \rightarrow i^{\prime}}$ of all 
the rate constants for collisions of $|M_N,p \rangle$ $=|+1,- \rangle$
molecules in which at least one molecule
relaxes into one of the strong-field-seeking states $||M_N|,p \rangle$
$= |0,- \rangle$, $|0,+ \rangle$, or $|1,+ \rangle$ (see Fig. 1).
Again both the ${\cal E}=0$ V/cm and ${\cal E}=200$ V/cm cases are shown, 
and the rates are separated into $L=0$ and $L=2$ initial states.  
These relaxation rates are dominated by exothermic processes which
exhibit their own characteristic threshold behavior. 
The rates for $L=0$ partial waves become
independent of energy at threshold, while the behavior
of $L=2$ partial waves transforms from the usual Wigner result
$K_{\rm relax} \propto E^2$ at low field, to a $K_{\rm relax} \propto E$ behavior at
high field when the dipoles are activated \cite{Mott}.

More significantly, the values of the loss rate constant are substantially 
boosted by the presence of an electric field, even for s-wave collisions.  This 
is simply because the strong dipolar interactions that drive inelastic
transitions are also made stronger in an electric field.  Figure
4 shows the threshold relaxation rate as a function
of electric field.  Even at low field the rates are fairly large,
since dipole interactions are still present at small $R$.
As the field grows to ${\cal E}\sim 100$ V/cm, 
where the dipoles turn on, the rates rise sharply.  In the particular
model considered here the rates are boosted  nearly two
orders of magnitude by the field.

Weak-electric-field seeking states should quite generally suffer these large 
loss rates.  Quantum mechanically this follows from the fact that the 
large-$R$ Hamiltonian ${\hat H}_{\rm fs} + {\hat H}_{\rm field}$ is 
diagonal in the laboratory frame, while the interaction Hamiltonian 
${\hat H}_{\rm dip-dip}$ is diagonal in the body frame that
joins the centers-of-mass of the two molecules.  The former
is stronger at large separations $R$, while the latter dominates
at small $R$.  Molecules that start out in
eigenstates of ${\hat H}_{\rm fs} + {\hat H}_{\rm field}$ are
thus distributed over all the different states of ${\hat H}_{\rm dip-dip}$
during the collision, and re-assembled into an assortment of
eigenstates of ${\hat H}_{\rm fs} + {\hat H}_{\rm field}$ as
the molecules separate.  Since the asymptotic states are completely
deconstructed and reconstructed during this collision, in general
it is expected that the probability for inelastic scattering 
is roughly the same as that for elastic scattering, and that
therefore the rates are comparable.

This is of course the same kind of physics that governs spin-exchange 
collisions in the alkali atoms, which are driven by the
competition between hyperfine-plus-magnetic field interactions
at large $R$, and exchange potentials at small $R$ \cite{Stoof}.  In the case
of alkali atoms it is possible that the short-range phase shifts,
from singlet and triplet total electronic spin states, can interfere
in such a way as to eliminate probabilities for inelastic
processes \cite{Burke}.  For molecules this coincidence seems
unlikely, however, since there are are many degrees of freedom
at short range, all of which would have to contribute nearly identical
scattering phase shifts in order cancel loss rates.

In conclusion, dipolar molecules electrostatically trapped
in weak-field-seeking states are
succeptible to state-changing collisions that can rapidly deplete
the trapped gas.  Moreover, these rates can grow in the presence of
the trapping electric field, which effectively turns on the
full dipolar coupling at large intermolecular separation.
Although this conclusion has been demonstrated using a particular toy 
model, the physics is quite general and should apply 
to any polar species.  It is therefore recommended that dipolar 
molecules be trapped in strong-field-seeking states, where inelastic 
channels are absent at low temperatures.  This kind of trapping cannot
be achieved in a static trap, but would require a time-varying
electric field.  Magnetic dipoles in strong-field seeking states have
indeed been confined in such traps, using either time-varying fields
\cite{Cornell} or microwave cavities \cite{Phillips}.
A more conventional magnetic trap may also be useful,
although the influence of the electric dipoles on losses in
magnetic traps would have to be explored.

More broadly, an externally applied electric field is seen to have
a profound influence on the collision dynamics
of ultracold polar molecules, even to the extent of altering the
threshold behavior.  Preliminary results on the
properties of quantum degenerate gases with dipolar interactions have
been reported in the literature \cite{Yi,Pfau}.  More detailed scattering
calculations are  required to help shape the study of
these unusual substances \cite{Bohnfuture}.

This work was supported by the National Science Foundation.
I acknowledge useful discussions with E. Cornell and C. Greene.

\epsfxsize = 3in
\epsfbox{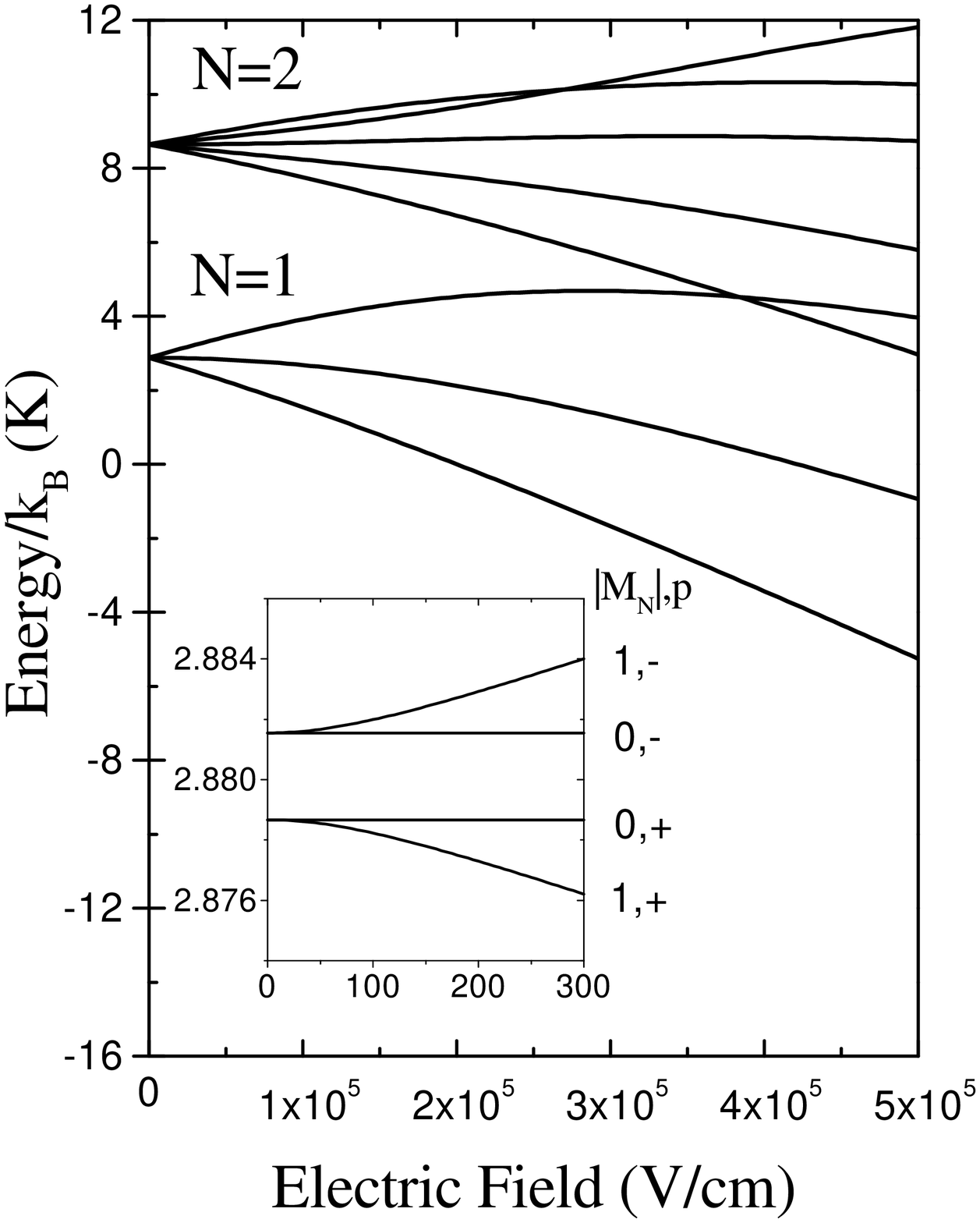}
\begin{figure}
\caption{Stark energy levels for the model molecules, which 
have $^1\Pi$ electronic symmetry.  This paper concentrates on collisions
between molecules in their $|M_N,p \rangle$ $=|+1,- \rangle$
states, which are weak-field seekers (see inset).}
\end{figure}

\epsfxsize = 3in
\epsfbox{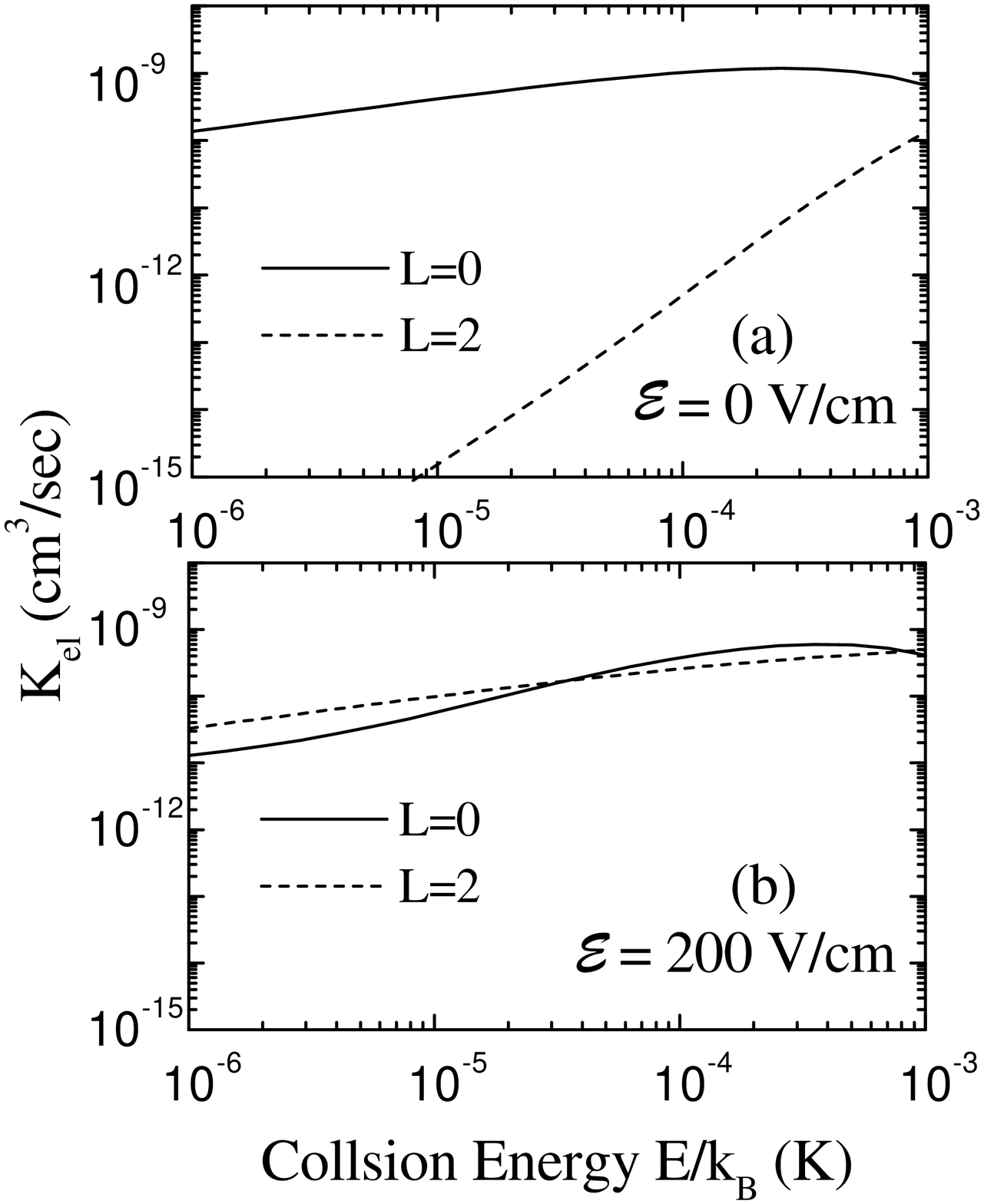}
\begin{figure}
\caption{Elastic scattering rate constants for $|M_N,p \rangle$ 
$=|+1,- \rangle$ molecules.
Shown are results for low-field [${\cal E}=0$ V/cm in (a)] and
high-field [${\cal E}=200$ V/cm in (b)] regimes.  Notice that
in the presence of an electric field the d-wave ($L=2$) rate is significantly
boosted, and acquires a $K_{\rm el} \propto E^{1/2}$ threshold behavior.}
\end{figure}

\epsfxsize = 3in
\epsfbox{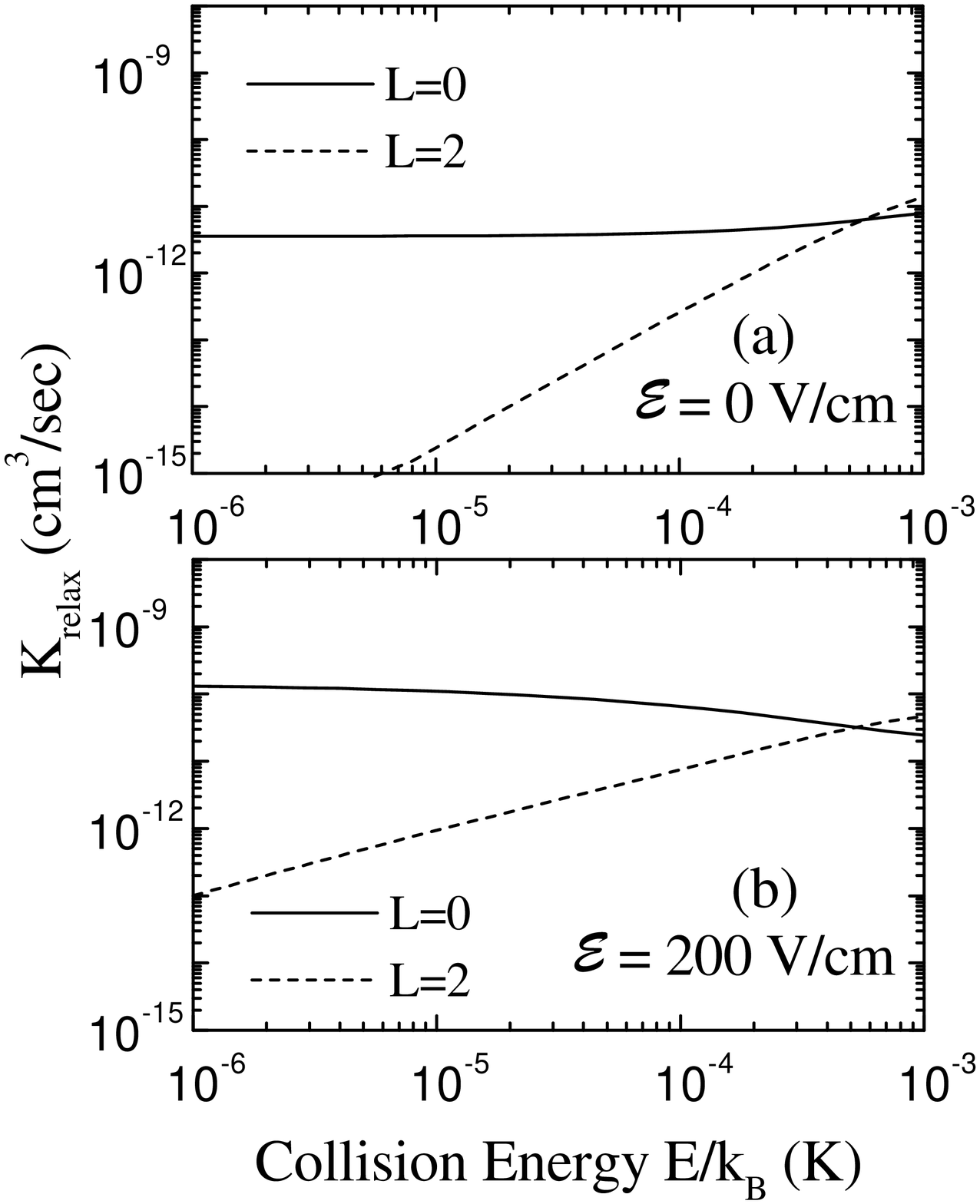}
\begin{figure}
\caption{Dipolar relaxation rate constants for $|M_N,p \rangle$ 
$=|+1,- \rangle$ molecules,
at the same field values as in Fig. 2.  The rates are substantially higher
in the presence of an electric field, as explained in the text.}
\end{figure}

\epsfxsize = 3in
\epsfbox{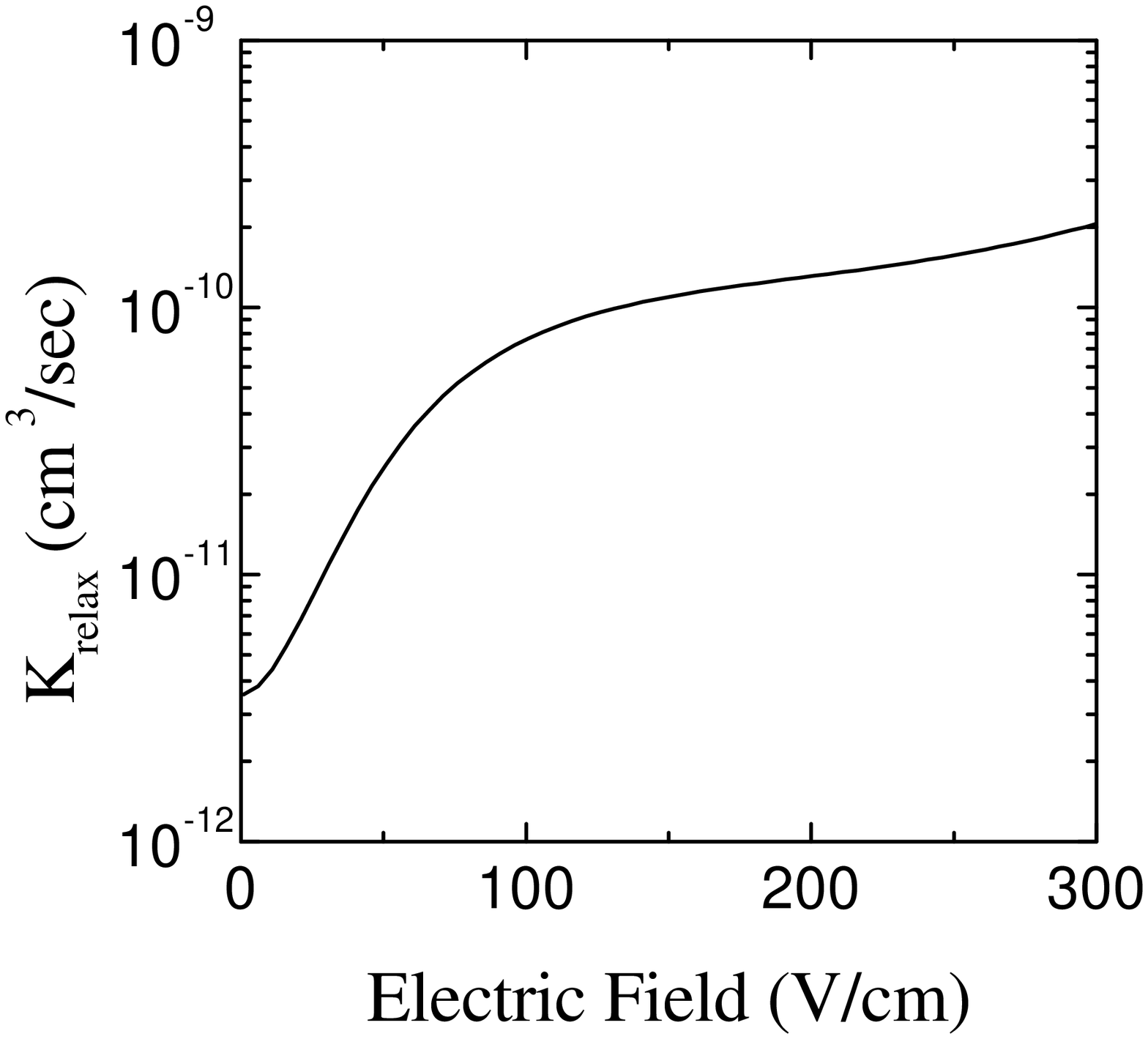}
\begin{figure}
\caption{Electric field dependence of the total relaxation rates for molecules
initially in their $|M_N,p \rangle$ $=|+1,- \rangle$ state, in the zero-collision-energy
limit.  These rates rise sharply by
${\cal E}=100$ V/cm, where the  molecular dipoles are ``activated''
by a suitable admixture of parity eigenstates.}
\end{figure}

\end{document}